\documentclass{pasj00}
\draft
\title{Detection of an H$\alpha$ Emission Line on a Quasar, RX J1759.4+6638, at $z=4.3$ with AKARI}

\author{Shinki \textsc{Oyabu}\altaffilmark{1,*},
  Takehiko \textsc{Wada}\altaffilmark{1},  Youichi
  \textsc{Ohyama}\altaffilmark{1}, 
  Hideo \textsc{Matsuhara}\altaffilmark{1}, \\
  Toshinobu \textsc{Takagi}\altaffilmark{1}, 
  Takao \textsc{Nakagawa}\altaffilmark{1},  
  Takashi \textsc{Onaka}\altaffilmark{2}, 
  Naofumi \textsc{Fujishiro}\altaffilmark{1}, \\
  Daisuke \textsc{Ishihara}\altaffilmark{2},
  Yoshifusa
  \textsc{Ita}\altaffilmark{1}, Hirokazu \textsc{Kataza}\altaffilmark{1},
  Woojung \textsc{Kim}\altaffilmark{1}, \\
  Toshio \textsc{Matsumoto}\altaffilmark{1}, 
  Hiroshi \textsc{Murakami}\altaffilmark{1}, 
  Itsuki \textsc{Sakon}\altaffilmark{2},
  Toshihiko \textsc{Tanab\'{e}}\altaffilmark{3},\\ 
  Kazunori \textsc{Uemizu}\altaffilmark{1}, 
  Munetaka \textsc{Ueno}\altaffilmark{4},  
  Fumihiko \textsc{Usui}\altaffilmark{1},
  Hidenori \textsc{Watarai}\altaffilmark{5},\\
  and Kanae  \textsc{Haze}\altaffilmark{1}}

\altaffiltext{*}{e-mail: oyabu@ir.isas.jaxa.jp}
\altaffiltext{1}{Institute of Space and Astronautical Science, Japan
  Aerospace Exploration Agency, \\
  3-1-1 Yoshinodai, Sagamihara, Kanagawa
229-8510, Japan}
\altaffiltext{2}{Department of Astronomy, School of Science,
  University of Tokyo, \\
  Bunkyo-ku, Tokyo 113-0033, Japan}
\altaffiltext{3}{Institute of Astronomy, University of Tokyo, \\
  2-21-1
  Osawa, Mitaka, Tokyo, 181-0015, Japan}
\altaffiltext{4}{Department of Earth Science and Astronomy, University
  of Tokyo, \\
  3-8-1, Komaba, Megro-ku, Tokyo 153-8902, Japan}
\altaffiltext{5}{Office of Space Applications, Japan Aerospace
  Exploration Agency, \\
  Tsukuba, Ibaraki, 305-8505, Japan}

\Received{---}
\Accepted{---}
\SetRunningHead{Oyabu et al.}{H$\alpha$ emission line detection of a
  quasar at $z=4.3$.}
\begin{document}
\maketitle

\begin{abstract}
We report the detection of an H$\alpha$ emission line in the low
resolution spectrum of a quasar,
RX J1759.4+6638, at a redshift of 4.3 with the InfraRed Camera (IRC)
onboard the AKARI. This is the first spectroscopic detection of an
H$\alpha$ emission line in a quasar beyond z=4. 
The overall spectral energy distribution (SED) of RX J1759.4+6638 in
the near- and 
mid-infrared wavelengths agrees with a median SED of the nearby quasars and the
flux ratio of $F(\mathrm{Ly}\alpha)/F(\mathrm{H}\alpha)$ is
consistent  
with those of previous reports for lower-redshift quasars.
\end{abstract}

\section{Introduction}

The broad emission lines of high-redshift quasars are luminous enough
to study physical properties of quasars such as
central black hole mass, accretion rate and metallicity of broad
emission line
regions in the early universe. 
Recent optical observations of emission lines such as C\emissiontype{IV} and
N\emissiontype{V} reveal 
supersolar abundances in quasar broad emission line regions even at
$z>4$ 
\citep{dietrich03}. 
The lack of evolution of the Fe\emissiontype{II}/Mg\emissiontype{II}
UV emission line 
ratio of quasars apparently continues out to $z$=6.4
\citep{barth03,iwamuro04}.
The rest-frame optical
Fe\emissiontype{II} emission as well as hydrogen Balmer emission lines in
high-redshift quasars cannot be measured from the ground facilities. 

These measurements can be pursued with the AKARI \citep{murakami07} which
has a unique capability to take 
near-infrared spectra in $2-5$\micron~from the space as well as
mid-infrared (5-14, 18-26\micron) with the InfraRed
Camera (IRC; \cite{onaka07,ohyama07}). Therefore one can trace the
redshifted emission lines toward the high-redshift universe with the AKARI.

A quasar, RX J1759.4+6638, at a redshift of 4.320 was discovered as the
most distant ROSAT X-ray selected object known at the time by
\citet{henry94} near the North Ecliptic Pole (NEP). 
In this paper, we report the near-infrared spectroscopy of this quasar
and the detection of an H$\alpha$ emission line at
$z=4.3$. Previous studies \citep{espey89,nishihara97} performed
the measurements of the
H$\alpha$ emission line in high-redshift quasars in the near-infrared
spectroscopy 
and the redshifts of their sample only reached $z \sim
2.4$ because the wavelength of redshifted H$\alpha$ emission lines is
affected by strong thermal emission of a telescope. 
Therefore this is the first spectroscopic detection of an H$\alpha$ 
emission line in a quasar beyond z=4, while
H$\alpha$ detections in photometric observations are
reported by \citet{jiang06}
for quasars at z$\sim 6$ by using the Spitzer IRAC. 
Our spectroscopic detection is caused by a benefit of high sensitivity 
spectroscopy with a cooled telescope from the space.

Through this paper, we adopt a flat cosmology with $H_0=71\ \mathrm{km}\
\mathrm{s}^{-1}\ \mathrm{Mpc}^{-1}$, $\Omega=0.27$ and $\lambda =
0.73$ \citep{spergel03}.  

\section{Observation and Data Reduction}

RX J1759.4+6638 was observed by the AKARI IRC on 2006 April, May,
October and 2007 February. As summarized in Table \ref{tab:tab1},
we made six pointed
observations using the IRC spectroscopic mode AOT04. There are three
observing modes: NP, NG, and NPNG modes. 
The NP mode uses the   
near-infrared prism with spectral resolving power $R$ $\sim$ 19 at
3.5$\mu$m. In this mode,
a target is put on
the imaging area of 
near-infrared detector. In the NP mode, a MIR-S spectroscopic
observation of a target is performed 
simultaneously. The NG mode uses the near-infrared grism ($R \sim$ 120
at 3.6$\mu$m), in which 
a target is put on a small aperture for
a point source grism spectroscopy. The other, NPNG, is a
special observing mode for a calibration
using the prism and the grism in 
the first half observation and the second, respectively.
The first four observations with the NPNG mode were done for
the wavelength calibration using the planetary nebula NGC 6543 in the
performance verification phase.
During these observations, the spectra of RX J1759.4+6638 were
obtained serendipitously. The remaining observations were targeted
observations for this quasar using the AKARI Director Time in order to
confirm the serendipitous detections in the performance verification
phase. 

\begin{table}
    \centering
    \caption{Observing Log}\label{tab:tab1}
    \begin{tabular}{llc}
        \hline
        \hline
        Obs. ID & Obs. Date & Obs. mode$^*$\\
        \hline
        5020047.1 & 2006-04-29 & NPNG \\
        5020048.1 & 2006-04-29 & NPNG \\
        5020049.1 & 2006-05-02 & NPNG \\
        5020050.1 & 2006-05-02 & NPNG \\
        5124035.1 & 2006-10-10 & NG \\
        5124044.1 & 2007-02-09 & NP \\            
        \hline
        \multicolumn{2}{@{}l@{}}{\hbox to
          0pt{\parbox{60mm}{\footnotesize
              Note. 
              \par\noindent
              \footnotemark[$*$] 
              NP is a spectroscopy on the imaging area
              of detector with the near-infrared prism for
              resolution, R$\sim$20, and NG is a spectroscopy on the
             slit of ``Np'' position with R$\sim$80 of the
             near-infrared grism. 
             NPNG is prepared using prism
             and grism in the first half observation and the second,
             respectively.
            }\hss}}
    \end{tabular}
\end{table}

The data were processed through the IRC Spectroscopic
toolkit \citep{ohyama07} to produce calibrated data 
frames. The data were converted into dark-subtracted, linearity
corrected and flat-field corrected frames after data that had
strong cosmic rays on the target object were removed with visual
investigation. 
Multiple frames were combined, and one-dimensional background-subtracted
spectra of the target quasar were extracted from the
combined data with the aperture width of 3 pixels (4.5\arcsec in the
NIR). The 
resultant spectra were scaled by a factor of 
1.6 for the aperture correction. 
In addition to the error estimation by the IRC Spectroscopic toolkit
using the sky variation near a 
target, we also calculated the root mean square of signal between each
frames. The original error
estimation was recognized to be overestimated due to the contamination
of other sources and the variations of signals between each
frame were used as the background error of spectra here. 

The photometric calibration of the spectrum was based on
AKARI IRC observations of standard stars, while the 
wavelength calibration was based on observations of emission line
stars and planetary nebulae \citep{ohyama07}. 
At this time
we estimate that the overall uncertainty in the wavelength
calibration is $\sim0.05\mu$m at 3.5$\mu$m. 

\section{Result}

\begin{figure}
    \centering
    \FigureFile(80mm,80mm){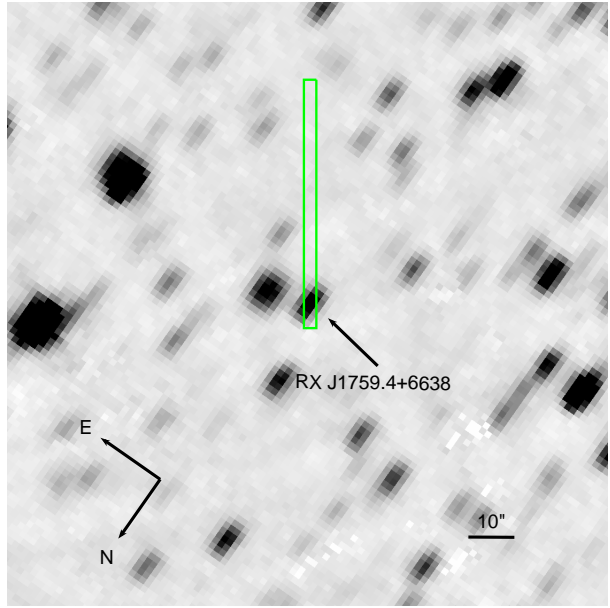}
    \caption{The image from the NEP-Deep in the
      N3-band\citep{oyabu07}. The image is rotated as the dispersion
      direction is upper at 2007 Feb. 9th as the spectrum position is
      shown with a green box. 
    }\label{fig:fig1} 
\end{figure}

The series of PV phase observations (ID 5020047.1, 5020048.1, 5020049.1
and 5020050.1) detected the H$\alpha$ emission
line only tentatively because the contamination of a faint
near-infrared 
source affected positions of the H$\alpha$ emission line. The targeted
NP
observation (ID 5124044.1) was arranged after we checked that the 
dispersion direction was free from any other object using the deep
N3-band image of the AKARI North Ecliptic Pole (NEP)
Survey (Figure \ref{fig:fig1}; \cite{oyabu07}). The 
NG spectroscopy (ID 5124035.1) and all MIR-S data, which were taken
simultaneously during NP observations,
provided spectra with low signal-to-noise ratio only.

\begin{figure}
    \centering
    \FigureFile(80mm,50mm){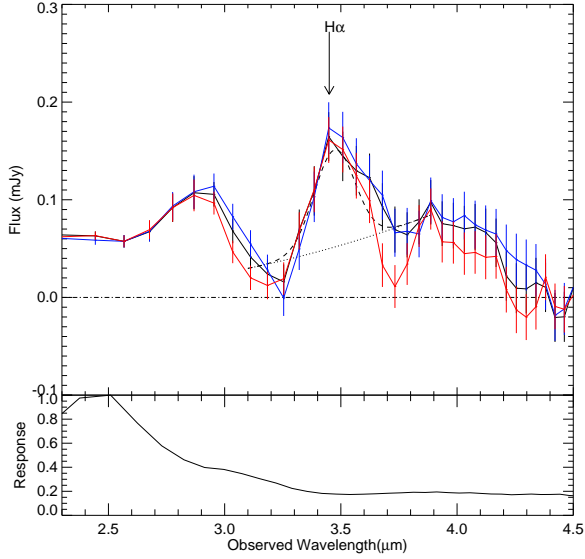}
    \caption{The NP spectrum of RX J1759.4+6638 (black solid line;
      top). Blue and red lines indicate the result from first and 
      second half of data, respectively. 
      The vertical lines are
      error bars consisting of background variation, wavelength
      calibration, flatten calibration and flux calibration errors. The
      dashed and 
      dotted lines show the fitted Gaussian and the assuming continuum,
      respectively. For reference, a normalized response curve of the NP as a
      function of wavelength (bottom). 
    }\label{fig:fig2}
\end{figure}

Figure \ref{fig:fig2} presents the resultant observed spectrum of the
NP observation on the top panel with a black thick line. 
On the spectrum, there are two features; a bump at the wavelength of
3.47 \micron~and 
a dip on the shorter wavelength side of the bump. 
The bump is located at the H$\alpha$ emission line corresponding to
$z$=4.29.
For the check of their reliabilities, we
divided their data in half and reduced them separately. On the top
panel of Figure \ref{fig:fig2}, two spectra of first and second half
data are also shown in blue and red lines, respectively. 
Both the dip
and the bump are shown in both data sets. In addition, we also checked
that the response function of the NP prism did not produce these features
as shown on the bottom panel of Figure \ref{fig:fig2}. Thus both
features, the bump and the dip, are real. The other possible
explanation of the bump is the  
contamination of other sources. However, as mentioned above, we
arranged the NP observation when the dispersion direction was free
from any other objects. 
%In Figure \ref{fig:fig1}, we present a finding chart of 
%this quasar in the N3-band.
There is no contamination in the dispersion direction as shown in
Figure \ref{fig:fig1}, 
and therefore we conclude that the detection of the H$\alpha$ emission
line is reliable. 

To determine the wavelength
center and the full width half maximum (FWHM), we fit a
Gaussian function to this spectrum after smoothing the spectrum with 3
pixels (Figure \ref{fig:fig2}). 
The detail measurements of
the H$\alpha$ emission line are summarized in Table \ref{tab:tab1}. 
The redshift from the H$\alpha$ 
emission line is found to be $\mathrm{z}=4.29\pm0.06$,
in agreement with the previous measurements of redshifts, $z$=4.320
\citep{henry94} and
4.32 \citep{constantin02} by using
restframe-ultraviolet emission lines from the ground-base optical
spectroscopy. The H$\alpha$ line flux is $4.9 \pm 1.1 \times 10
^{-22}\ 
\mathrm{W}\ \mathrm{cm}^{-2}$, corresponding to the luminosity of
H$\alpha$ emission line $L(\mathrm{H\alpha}) = 7.5 \pm 1.7 \times
10^{36}\ \mathrm{W}$ at $z$=4.3. While the H$\alpha$ emission
line flux is affected by two [N \emissiontype{II}] lines located at blue and
red side of the H$\alpha$ line, their contribution of [N \emissiontype{II}]
emission lines to the H$\alpha$ line flux is only 3 percents in the
Sloan Digital Sky Survey (SDSS) composite 
\citep{vanden01} and is ignored in this paper.  
The FWHM of the H$\alpha$ emission
line is found to be unresolved and $18000 \pm 4000 \mathrm{km}\
\mathrm{s}^{-1}$ which is 
comparable to the instrumental resolution $\sim 20000 \mathrm{km}\
\mathrm{s}^{-1}$ when we made the 3-pixel 
smoothing data.
The restframe equivalent width (EW) of the H$\alpha$ emission
line is
0.071$\pm$0.015 $\mu$m which is converted from the observed EW of
0.38$\pm$0.08 $\mu$m at $z$=4.3. 
However there is 
a dip of the continuum around 3.2$\mu\mathrm{m}$ in Figure
\ref{fig:fig2} and the dip makes it difficult to measure the
H$\alpha$ line flux and the EW accurately.  
This dip might be a broad absorption line feature, while there are no
features of broad absorption lines in the restframe UV
spectra\citep{henry94,constantin02}. Thus the reason of this
dip is still uncertain.

\begin{figure}
    \centering
    \FigureFile(80mm,50mm){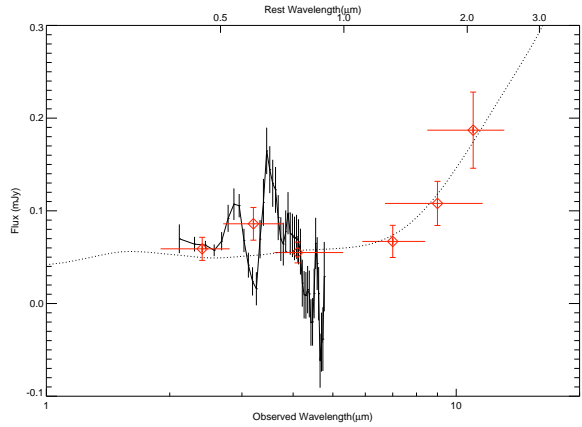}
    \caption{Comparison of the spectrum of RX J1759.4+6638 with the
      photometric study. The solid line is the IRC
      spectrum as shown 
      in Figure \ref{fig:fig1} with 1$\sigma$ error bars, while
      red diamonds presents the 
      photometric result from \citet{oyabu07}. The red horizontal and
      vertical bars on
      each point presents the band width and the 1$\sigma$ photometric
      error, 
      respectively. For comparison, the median
      spectral energy distribution
      of low-redshift quasars \citep{elvis94} is plotted in a dotted
      line after 
      scaled at the N4-band.
      In \citet{oyabu07},
      this quasar is called as CXOSEXSI
      J175928.1+663851.}\label{fig:fig3} 
\end{figure}

We also measured the NIR and MIR-S photometric fluxes of this quasar
with the IRC 
during the course of the AKARI NEP-Deep Survey 
\citep{oyabu07}\footnote{In \citet{oyabu07},
  this quasar is called as CXOSEXSI J175928.1+663851.}. 
The results of the NIR and MIR-S bands are summarized in Table
\ref{tab:tab3}. 
Figure \ref{fig:fig3} presents the comparison of the IRC spectrum of RX
J17759.4+6638 with IRC photometric result as well as the
median spectral energy distribution (SED) of low-redshift
quasars \citep{elvis94}. 
Assuming that the N3-band flux includes the H$\alpha$ emission line flux
and the N2- and N4-bands represent continuum level, the flux of the
H$\alpha$ emission line, $F(\mathrm{H}\alpha) = 
4.0 \pm 2.6\ \times 10^{-22}\ \mathrm{W}\ \mathrm{cm}^{-2}$, from the
photometric data agrees with the spectroscopic measurement of the H$\alpha$
emission line. The uncertainty of the line flux from the
photometric data is dominated by the
systematic errors of photometric calibration. 
The observed fluxes 
of emission lines and continuum level are consistent with the
photometric study within their uncertainties.
Comparing our measurements with the
median SED of low-redshift
quasars \citep{elvis94},
RX J1759.4+6638 has quite typical spectral
energy distribution, which suggests no significant evolution in the
rest-frame optical- and near-infrared wavelengths. 

\begin{longtable}{lccccc}
    \caption{Observed H$\alpha$ line measurement of RX J1759.4+6638}\label{tab:tab2}

    \hline
    \hline
    Line & Observed Wavelength & Redshift & Observed Flux$^*$ & FWHM$^{\dagger}$ &
    Observed EW$^{\star}$ \\
         &  $(\mu\mathrm{m})$  &   & $(10^{-22}\ \mathrm{W}\ \mathrm{cm}^{-2})$ & $(\mathrm{km}\
         \mathrm{s}^{-1})$ & $(\mu\mathrm{m})$ \\
    \hline
    \endfirsthead
    H$\alpha$ 0.6563$\mu$m &  3.47$\pm$0.05  & 4.29$\pm$0.06 &  4.9 $\pm$ 1.1  &
    18000 $\pm$ 4000$^{\ddagger}$  & 0.38$\pm$0.08 \\
    \hline
    \multicolumn{6}{@{}l@{}}{\hbox to
      0pt{\parbox{165mm}{\footnotesize
          Note. 
          \par\noindent
          \footnotemark[$*$]Flux is measured by direct
          integration of the line flux.\\
          \footnotemark[$\dagger$]The observed full width half maximum(FWHM) of
          emission line.\\
          \footnotemark[$\star$]The observed equivalent width(EW) of
          emission lines.\\
          \footnotemark[$\ddagger$]The observed FWHM is unresolved
          with the instrumental resolving power $\sim$
          20000 $\mathrm{km}\ \mathrm{s}^{-1}$ after 3 pixel
          smoothing at 3.5$\mu$m. 
        }\hss}}
\end{longtable}

\begin{table}
    \centering
    \caption{The AKARI/IRC photometric measurements of RX J1759.4+6638 from
      \citet{oyabu07}}\label{tab:tab3} 
    \begin{tabular}{lrr}
        \hline
        \hline
        Band($\lambda_{\mathrm{ref}}$) & Flux($\mu$Jy) & Flux error($\mu$Jy)$^*$\\
        \hline
        N2 (2.4\micron) & 59   & 4\\
        N3 (3.2\micron) & 86   & 4\\
        N4 (4.1\micron) & 55   & 2\\
        S7 (7.0\micron)  & 67   & 11\\
        S9W (9.0\micron) & 108 & 10 \\
        S11 (11.0\micron) & 187 & 17\\
        \hline
        \multicolumn{2}{@{}l@{}}{\hbox to
          0pt{\parbox{80mm}{\footnotesize
              Note. 
              \par\noindent
              \footnotemark[$*$] Only statistical errors of the
              fluxes are
              presented. About 20 percents of the systematic errors
              exist during the photometric calibrations. 
            }\hss}}
    \end{tabular}
\end{table}

\section{Discussion}

The flux ratio of the observed H$\alpha$ emission line to Ly$\alpha$ would
be useful to investigate the physical conditions in the
broad-line region and the reddening to this region. \citet{henry94}
observed the rest-frame ultraviolet spectra
from ground-based telescopes, and they measured the Ly$\alpha$
emission line flux of 
$F(\mathrm{Ly}\alpha) = 1.4 \times 10^{-21}\ \mathrm{W}\ 
\mathrm{cm}^{-2}$ on June 1993.  
\citet{constantin02} made a new measurement of $F(\mathrm{Ly}\alpha) =
6.4\times 10^{-22}\ \mathrm{W}\ \mathrm{cm}^{-2}$ on June 1999
with the Keck Telescope \footnote{Observation details are provided
  from A. Constantin in private communication.}.
The
Ly$\alpha$ line flux from \citet{constantin02} changed into less than
half of the measurements in \citet{henry94} on a timescale of about
1 yr (rest frame) suggesting that this quasar is variable.
We note that the X-ray observations of this quasar \citep{grupe06}
also reported variability.
In addition, the line profile of Ly$\alpha$
is strongly affected by the intervening column of intergalactic
neutral hydrogen \citep{henry94,constantin02}. 
These problems make
straightforward comparisons between H$\alpha$ and Ly$\alpha$ line
fluxes difficult. However assuming that the Ly$\alpha$ emission line
flux in \citet{constantin02} be minimum, the
lower limit, $F(\mathrm{Ly}\alpha)/F(\mathrm{H}\alpha)> 1.3$, is
calculated.
This ratio is consistent with other observational results;
$F(\mathrm{Ly}\alpha)/F(\mathrm{H}\alpha) = 3.2$ of the SDSS composite
quasar \citep{vanden01} and
$F(\mathrm{Ly}\alpha)/F(\mathrm{H}\alpha) = 1.2 - 7.4$ of the
low-redshift quasars at z=0.061 - 0.555 \citep{tsuzuki06}, although 
the theoretical values of
$F(\mathrm{Ly}\alpha)/F(\mathrm{H}\alpha) = 10 - 12$ from pure
recombination calculation \citep{osterbrock06} are greater than
observations. 

AKARI will extend the spectroscopic sample of high-redshift quasars'
spectroscopy 
for more detailed studies of H$\alpha$ emission lines at $z>4$. 
Specifically, one of the AKARI Mission Programs, ``The unbiased
Slit-less 
spectroscoPIC surveY of galaxies (SPICY)'' \citep{matsuhara06}, is
designed and conducted to make 
$\sim$ 0.5 square degree scale survey with slit-less spectroscopy with
AKARI 
IRC wavelength and will provide a new sample of H$\alpha$ emission lines
in the high-redshift universe. 

\section{Summary}

We present the detection of an H$\alpha$ emission line on the low
resolution spectrum of a 
quasar RX J1759.4+6638 at a redshift of 4.3 with the IRC onboard the
AKARI after careful
consideration of possible artifact and contamination.  
This is the first spectroscopic detection of an
H$\alpha$ emission line in a quasar beyond z=4. 
Our spectroscopic measurement shows a good agreement with the
photometric data from 
the AKARI NEP-Deep Survey within their uncertainties. 
The overall SED of RX J1759.4+6638 in the near- and
mid-infrared wavelengths also agrees with a median SED of the nearby 
quasars. The
flux ratio of $F(\mathrm{Ly}\alpha)/F(\mathrm{H}\alpha)$ of this
quasar is consistent 
with those of previous report for lower-redshift quasars.
These results
suggest no significant evolution of the quasar's features at $z=4.3$
in the optical- and near-infrared wavelength (rest frame). 

\vspace{1cm}
AKARI is a JAXA project with the participation of ESA. We thank all
the members of the AKARI project for their continuous 
help and support.

\end{document}